\documentclass[11pt]{article}
\usepackage{cospar}
\usepackage{url}

\def\lesssim{\lower4pt\hbox{${\buildrel < \over \sim}$}}
\def\gtrsim{\lower4pt\hbox{${\buildrel > \over \sim}$}}

\usepackage{graphicx}
\usepackage[figuresright]{rotating}

\title{X-RAY LINES AND ABSORPTION EDGES IN GRBS AND THEIR AFTERGLOWS}

\author{M. B\"ottcher\address{Department of Physics and Astronomy,
Ohio University, Athens, OH 45701, USA}} 

\begin{document}

\maketitle

\begin{abstract}
Absorption and Reprocessing of Gamma-ray burst radiation 
in the environment of cosmological GRBs can be used as a
powerful probe of the elusive nature of their progenitors.
In particular, transient X-ray emission line and absorption
features in the prompt and early afterglows of GRBs are
sensitive to details of the location and density structure
of the reprocessing and/or absorbing material. To date, there
have been only rather few detections of such features, and the 
significance is marginal in most individual cases. However, 
transient X-ray emission lines in GRB afterglows have now 
been found by four different X-ray satellites, which may 
justify a more detailed theoretical investigation of their 
origin. In this paper, I will first present a brief review of 
the status of observations of transient X-ray emission line and 
absorption features. I will then discuss general physics
constraints which those results impose on isotropy, homogeneity,
and location of the reprocessing material with respect to the
GRB source, and review the various currently discussed,
specific models of GRBs and their environments in which the 
required conditions could arise.

\end{abstract}

\section*{SUMMARY OF OBSERVED X-RAY LINES AND ABSORPTION FEATURES}

The precise localization of gamma-ray bursts (GRBs) by the BeppoSAX 
satellite, launched in 1996, has facilitated the subsequent discovery 
of X-ray and optical GRB afterglows, the measurement of redshifts of
GRBs and the firm establishment of their cosmological distance scale 
(at least for long GRBs with durations $t_{90} \gtrsim 2$~s) beyond 
any reasonable doubt. The physics of the radio through X-ray continuum 
afterglow emission are now believed to be rather well understood in
terms of the external synchrotron shock model (for a recent review
see, e.g., M\'esz\'aros 2002 or Dermer 2002). However, in spite of 
these significant advances, the ultimate source of GRBs is still a 
matter of vital debate. This is mainly due to the fact that the 
continuum GRB afterglows are the ``smoking gun'' of the GRB explosion, 
revealing only very little information about the initial energy source. 

However, even without a direct observation of the central engines of 
GRBs, it might be possible to infer their nature indirectly if detailed 
probes of the structure and composition of their immediate vicinity 
can be found. A promising candidate for such a probe is high-quality, 
time resolved X-ray spectroscopy of GRB afterglows, which is now 
becoming available with the new generation of X-ray telescopes like
{\it Chandra} and {\it XMM-Newton}, and the planned Swift mission,
scheduled for launch in September 2003. Previous and currently
operating X-ray telescopes have so far (status: November 2002) 
revealed marginal evidence for X-ray emission lines (mostly consistent 
with Fe~K$\alpha$ fluorescence lines) or radiative recombination 
edges in 5 GRBs, and one case of a transient X-ray absorption feature 
at an energy consistent with an iron K absorption edge. 

In the remainder of this section, I will give a brief review of the 
observed X-ray spectral features in GRBs, before I turn to the
implications of those observations in general terms. This will be
followed by a summary of the currently discussed GRB models which
could give rise to the conditions required to produce the observed
spectral features.

\subsection*{X-ray emission lines in GRB afterglows}

Marginal evidence for transient X-ray emission line features in
GRB afterglows have been reported for 5 Gamma-ray bursts: GRB~970508
(Piro et al. 1999), GRB~970828 (Yoshida et al. 2000), GRB~991216
(Piro et al. 2000), GRB~000214 (Antonelli et al. 2000), and 011211
(Reeves et al. 2002a). Tab. \ref{line_table} lists the key observational
features of these five GRB line detections: (1) the GRB identification,
(2) the redshift, (3) the time interval of the respective line 
detection, (4) the isotropic luminosity of the emission line, (5)
the instrument with which the detection was made, (6) additional
remarks concerning the line detection, and (7) the reference to
the detection and important contributions through re-analysis of 
the respective observations. Note that the beginning of the time
intervals listed in column (3) are generally set by the beginning
of the respective observations and might thus not be representative
of the onset of the line emission.

\begin{table}[h]
\vspace{-8mm}
\caption{Observed X-ray emission line features in GRB afterglows. 
$t_{\rm line}$ is the time interval after the GRB trigger during
which the respective feature(s) has (have) been detected. $L_{\rm line}$
is the luminosity in the respective emission line, assuming that the
line emission is isotropic.}
\label{line_table}
\begin{tabular}{ccccccc}
\hline
 GRB & $z$ & $t_{\rm line} {\rm [s]}$ & $L_{\rm line} {\rm [ergs \; s}^{-1}{\rm ]}$ & 
Instrument & Remarks & Reference(s) \\
\hline
\hline
970508 & 0.835 & $2 \cdot 10^4$     & $6 \cdot 10^{44}$ & BeppoSAX & Sec. outburst   & Piro et al. \\
       &       & $- 5.6 \cdot 10^4$ &                  &          & at time of line & (1999)       \\
       &       &                  &                  &          & detection       &                \\
       &       &                  &                  &          & Line width in-  & Paehrels       \\ 
       &       &                  &                  &          & consistent with & et al. (2000)  \\
       &       &                  &                  &          & photoionization &                \\
\hline
970828 & 0.958 & $1.2 \cdot 10^5$   & $5 \cdot 10^{44}$ & ASCA     & Rad. rec. cont. & Yonetoku    \\
       &       & -- $1.4 \cdot 10^5$ &                 &          & without line    & et al. (2001) \\
       & (0.33) &                 & ($5 \cdot 10^{43}$) & & ($\longleftarrow$ if em. & Yoshida \\
       &       &                  &                 &          & feature is in-  & et al. (1999)  \\
       &       &                  &                 &          & terpreted as    &                \\
       &       &                  &                 &          & Fe K$\alpha$ line) &             \\
\hline
991216 & 1.00  & $1.3 \cdot 10^5$   & $8 \cdot 10^{44}$ & Chandra  & Broad line +    & Piro et al. \\
       &       & -- $1.5 \cdot 10^5$ &                 & ACIS-S   & Rad. Rec. Cont. & (2000)      \\
       &       &                  &                 & + HETG   & $\sigma_L \sim 0.23$~keV &     \\
       &       &                  &                 &          & $\Longrightarrow$ high &      \\
       &       &                  &                 &          & velocity outflow &            \\
       &       &                  &                 &          & $v \sim 0.1$~c &             \\
\hline
000214 & 0.47  & $4 \cdot 10^4$     & $4 \cdot 10^{43}$ & BeppoSAX & No optical      & Antonelli   \\
       &       & -- $1.5 \cdot 10^5$ &                 &          & transient;      & et al. (2000) \\
       &       &                  &                 &          & $z$ based on    &             \\
       &       &                  &                 &          & X-ray line      &             \\
\hline
011211 & 2.14  & $4.0 \cdot 10^4$   & Si XIV: $6.4 \cdot 10^{44}$
                                                    & XMM-     & No iron line;   & Reeves et al. \\
       &       & -- $6.7 \cdot 10^4$ & S XVI: $6.2 \cdot 10^{44}$
                                                    & Newton   & reality of line & (2002a,b)     \\
       &       &                  & Ar XVIII: $4.4 \cdot 10^{44}$
                                                    &          & detection con-  & Borozdin \& Tru-  \\
       &       &                  & Ca XX: $2.5 \cdot 10^{44}$
                                                    &          & troversial      & dolyubov (2002)  \\

\hline
\end{tabular}
\end{table}

In the {\it BeppoSAX} NFI observation of the afterglow of GRB~970508,
Piro et al. (1999) detected evidence, at the $\sim 3 \, \sigma$ level,
for an emission line feature consistent with a 6.7~keV Fe~K$\alpha$ 
line from highly ionized iron, at the redshift of the burst at 
$z = 0.835$. The X-ray afterglow of this burst exhibited a secondary
outburst after $\sim 6 \times 10^4$~s, and the disappearance of the
line from the X-ray spectrum seemed to be coincident with the onset of
this secondary X-ray outburst (Piro et al. 1999). Paehrels et al. (2000) 
re-analyzed the NFI spectrum containing the line. Fitting the spectrum
with a plasma emission model in photoionization equilibrium, assuming
an illuminating continuum identical to the GRB afterglow emission, 
they found that such a fit would either require too large a redshift,
inconsistent with the redshift of the burst, or too high a temperature,
inconsistent with the a priori assumption of the ionization state of
the line-emitting medium being dominated by photoionization. Paehrels
et al. (2000) interpreted this inconsistency as possible evidence for 
thermal plasma emission rather than emission from a photoionized plasma.

Interestingly, the second GRB for which evidence for an X-ray emission
feature in the redshifted Fe~K$\alpha$ energy regime was found, appears
to be opposite in that respect. Before the redshift of GRB~970828 
could be determined, Yoshida et al. inferred a redshift of $z = 0.33$ 
from the energy of the line feature in the {\it ASCA} spectrum of the
afterglow of this GRB, assuming that it corresponds to an Fe~K$\alpha$ 
line at 6.4~keV in the burst rest frame. In a later re-analysis, after 
the likely host-galaxy identification of GRB~970828, associated with a 
redshift of $z = 0.958$ (Djorgovski et al. 2001), Yoshida et al. (2001) 
and Yonetoku et al. (2001) argued that the emission feature is consistent
with this redshift if it is a radiative recombination continuum (RRC) edge 
rather than a fluorescence or recombination line. They argue that the
electron temperature in the plasma responsible for the emission feature
is inconsistent with the ionization temperature in thermal equilibrium, 
and would therefore indicate photoionization as the dominant ionization
mechanism.

Marginal evidence for a RRC was also reported for the {\it Chandra} 
ACIS-S+HETG spectrum of the afterglow of GRB~991216 (Piro et al. 2000). 
Piro et al. (2000) interpret the width and the apparently blue-shifted 
best-fit energy of the iron line with respect to the redshift of the 
host galaxy at $z = 1.00$ as the signature of a directed outflow velocity 
of $v \sim 0.1 \, c$ of the source of the recombination line and 
the RRC. If this interpretation is correct, it might be indicative 
of a supernova explosion a few months prior to the GRB. 

For GRB~000214, no optical counterpart could be identified. In this
case, the only information concerning its redshift comes from the
X-ray emission line detected by the {\it BeppoSAX} NFI, and was
estimated to be $z = 0.47$, if the emission line at $E = (4.7 \pm
0.2)$~keV is interpreted as the redshifted Fe~K$\alpha$ line from
hydrogen-like iron (Antonellii et al. 2000).

Recently, Reeves et al. (2002a,b) reported the marginal detection
of a set of emission lines in the {\it XMM-Newton} spectrum of the
early afterglow of GRB~011211. If real, these features would be
peculiar in that they show evidence for K$\alpha$ lines from the 
hydrogen-like ions of lighter elements, such as Si~XIV, S~XVI, 
Ar~XVIII, and CA~XX, but no indication of an Fe~K$\alpha$ line 
or RRC edge. Those lines appear to be blue-shifted by an average
of $v \sim 0.1 \, c$ with respect to the likely redshift of the
burst, $z = 2.14$. Reeves et al. (2002b) find that the addition
of the low-Z metal line system to a pure power-law fit to the
{\it XMM-Newton} EPIC-pn spectrum improves the $\chi^2/d.o.f.$
from $51.5/28$ to $21.9/22$. However, Borozdin \& Trudolyubov (2002)
have pointed out several caveats of the tentative line detections
in GRB~011211. (1) The set of low-Z metal emission lines appeared
only in the first $\sim 5$~ksec, prior to a re-pointing of the
{\it XMM-Newton} spacecraft. During that time, the source was
located very close to the edge of the CCD chip of the EPIC-pn
detector. Repeating the analysis of Reeves et al. (2002a), but
excluding 22~\% of the counts from the immediate vicinity of the
chip edge, they found that the significance of the lines dropped
disproportionally. (2) Evidence for the lines appears only in the 
EPIC-pn detector, with no such indication in the EPIC-MOS detectors. 
(3) The measured centroid energy of the most prominent line at 
$E = 0.70 \pm 0.02$~keV (attributed to Si~XIV by Reeves et al. 2002b), 
is coincident with a line in the background from a region on the chip
in the immediate vicinity of the edge. (4) In their analysis, Borozdin 
\& Trudolyubov (2002) find the {\it XMM-Newton} spectrum of GRB~011211
consistent with a pure power law, with no evidence for spectral evolution
in consecutive time intervals. The addition of an emission line system 
lowers the reduced $\chi^2$ from 1.03 to 0.82 (with 8 additional free
parameters), which might be an indication of over-fitting, rather than 
a statistically significant improvement of the fit. 

\subsection*{The Transient Absorption Feature in GRB~990705}

Atomic X-ray absorption features in GRB afterglows are rather common.
However, in order to distinguish such features from foreground
absorption and to associate them physically with the GRB, one 
needs to find evidence for variability of those absorption features 
on the GRB / early afterglow time scale (Perna \& Loeb 1998, 
B\"ottcher et al. 1999). Until now, only one such case has 
been observed: GRB~990705 (Amati et al. 2000). In this burst, a
transient absorption feature has been found at $(3.8 \pm 0.3)$~keV, 
consistent with an Fe~K absorption edge from neutral iron at $z = 0.86 
\pm 0.17$ at the redshift of the host galaxy at $z = 0.84$ (Lazzati 
et al. 2001). This absorption edge was only seen in the first 
13~s of the GRB, while no evidence for excess absorption was found 
in later segments of the {\it BeppoSAX} WFC observations. The best
fit to the segment with the most significant detection of the
absorption edge (6 -- 13~sec.), assuming an underlying power-law
absorbed by a neutral absorbing column $N_H$ with solar abundances
and an additional absorption edge yielded a depth of the absorption 
feature of $\tau = 1.4 \pm 0.4$ and $N_H = (3.5 \pm 1.4) \times
10^{22}$~cm$^{-2}$. Leaving the iron abundance in the neutral absorber 
as a free parameter in a power-law fit to the spectrum in this time 
segment resulted in $N_H = (1.32 \pm 0.3) \times 10^{22}$~cm$^{-2}$
and a relative overabundance of iron with respect to solar abundances
of $X_{\rm Fe} = 75 \pm 19$. Implications and possible interpretations
of these results will be discussed in the final section of this review.

\section*{GENERAL CONSTRAINTS FROM EMISSION LINE FEATURES}

Analytic estimates of the fluorescence and/or recombination line 
emission in photoionized media in GRB environments have been presented 
by many authors (e.g., M\'esz\'aros \& Rees 1998, Ghisellini et al. 1999,
Lazzati et al. 1999, B\"ottcher 2000). A very general estimate of 
the amount of iron required to produce the observed iron line features
can be derived from considering that in the course of complete 
photoionization of iron, on average $\sim 5$ Fe~K$\alpha$ photons 
are emitted, since it takes on average $\sim 12$ X-ray photons to 
ionize an initially neutral iron atom completely (taking into account 
the Auger effect), and the K$\alpha$ fluorescence yield is in the range 
0.3 -- 0.4 for the various ionization stages of iron. In dilute media,
if recombination is negligible, it would then take $N_{\rm Fe} \sim
2 \times 10^{56} L_{44} \, t_5$ iron atoms to produce an iron line of 
isotropic luminosity $L_{\rm line} = 10^{44} \, L_{44}$~ergs~s$^{-1}$ 
over a time scale of $\Delta t = 10^5 \, t_5$~s. If recombination and 
multiple ionization of the same iron atom enhance the efficiency of line 
production, one can introduce an ehnancement factor $f$, counting the number 
of times that a single iron atom can effectively contribute 5 K$\alpha$ 
photons in the process of repeated cycles of ionization and recombination. 
One then arrives at a general estimate of

\begin{equation}
M_{\rm Fe} \approx 0.16 \, {L_{44} \, t_5 \over f} \; M_{\odot}
\label{M_Fe_general} 
\end{equation}
In dilute, extended media, where recombination is inefficient, $f \sim 1$,
while in dense media with efficient recombination, $f \gg 1$. Due to light
travel time effects, the duration of the line emission will be at least
$t_{\rm line} > R \, (1 + z) \, (1 - \cos\theta_{\rm obs}) /c$, where $R$ 
is the extent of the reprocessing material. Thus, the size limit, together 
with the restriction that in the GRBs with line detections, there is no 
indication for excess X-ray absorption, leads to typical mass estimates 
of $M_{\rm Fe} \sim 0.1$ -- 1~$M_{\odot}$ of iron, confined in $R \lesssim 
10^{-3}$~pc, if the line emission were to originate in a dilute, 
quasi-isotropic environment. This is unlikely to be realized in any 
astrophysical setting, and may thus be ruled out (Ghisellini et al.
1998, B\"ottcher et al. 1999). We therefore find that inhomogeneous 
media with significant density enhancement outside our line of sight 
toward the GRB source are required (e.g., Lazzati et al. 1999, 
B\"ottcher 2000).

At this point, several geometries and general scenarios will have to be 
distinguished. First, the duration $t_{\rm line}$ of the line detection
can be set either by the light travel time effect, which puts the
reprocessor at distances of $10^{15} \, {\rm cm} \lesssim R \lesssim
10^{16}$~cm. These types of configurations are referred to as {\it
distant reprocessor models}. Alternatively, the duration of the line
emission can be determined by the duration of the illumination by a
more persistent, gradually decaying central source. In that case, the
reprocessor can be located much closer to the source than in the case
of the distant reprocessor models. For this reason, these scenarios
are referred to as {\it nearby reprocessor models} (see fig. 1). 
Second, we need to distinguish between different mechanisms ionizing 
the line-emitting iron atoms. The scenarios considered above are 
generally based on photoionization being the dominant ionization 
mechanism. However, it is very well conceivable that the material 
in the vicinity of GRBs is energized by shocks associated with the 
GRB explosion, and heated to temperatures of $T \sim 10^7$ -- $10^8$~K 
(e.g., Vietri et al. 1999). In this case, and for sufficiently high 
densities, collisional ionization may dominate over photoionization, 
and the iron K$\alpha$ line emission will be dominated by the 
recombination lines of H and He-like iron, accompanied by a radiative 
recombination continuum. We will refer to these models as ``thermal 
models'', as opposed to the ``photoionization models'' mentioned above. 

\section*{PHOTOIONIZATION MODELS}

General parameter studies on photoionization models have been done
by many authors. An important complication keep in mind is the
fact that the continuum illuminating the reprocessing material
might not be identical or even similar to the observed afterglow 
emission. Most importantly, if the reprocessor is located at a
distance of $R = 10^{15} \, R_{15}$~cm, and illuminated by GRB
(and early afterglow) emission from a relativistic blast wave
advancing at a coasting speed corresponding to $\Gamma = 100 \,
\Gamma_2$, then it will be swept up by the blast wave after a
typical illumination time of

\begin{equation}
t_{\rm ill} \sim 2 \, {R_{15} \over \Gamma_2^2} \; {\rm s}
\label{t_ill}
\end{equation}
unless the blast wave deceleration radius $r_{\rm dec} = 5 \times
10^{16} \, (E_{52} / n_0 \, \Gamma_2^2)^{1/3}$~cm is much less than
$10^{15}$~cm. Here, $E_{52}$ is the isotropic equivalent energy of
the GRB explosion in units of $10^{52}$~ergs~s$^{-1}$ and $n_0$ is 
the density of the (homogeneous) surrounding medium in units of 
cm$^{-3}$. The condition $r_{\rm dec} \ll 10^{15}$~cm would require
a very low explosion energy directed toward the reprocessor, a large
density of the decelerating external medium, and/or a very high bulk
Lorentz factor $\Gamma$, i. e. a very low baryon contamination.

General studies of the reprocessing efficiency and spectral and
temporal signatures of reprocessor models for X-ray emission lines
in GRB environments (e.g., Ballantyne \& Ramirez-Ruiz 2001, Lazzati 
et al. 2002, Ghisellini et al. 2002, Kallman et al. 2002) have 
yielded very useful insight into the general properties of such 
reprocessing features. The above caveat needs to be taken into 
account very carefully when applying these results to real GRBs 
and scaling illuminating spectra, ionization parameters etc., to 
properties derived from observed continuum afterglows of GRBs. 

\begin{figure}
\begin{center}
\includegraphics[width=14cm]{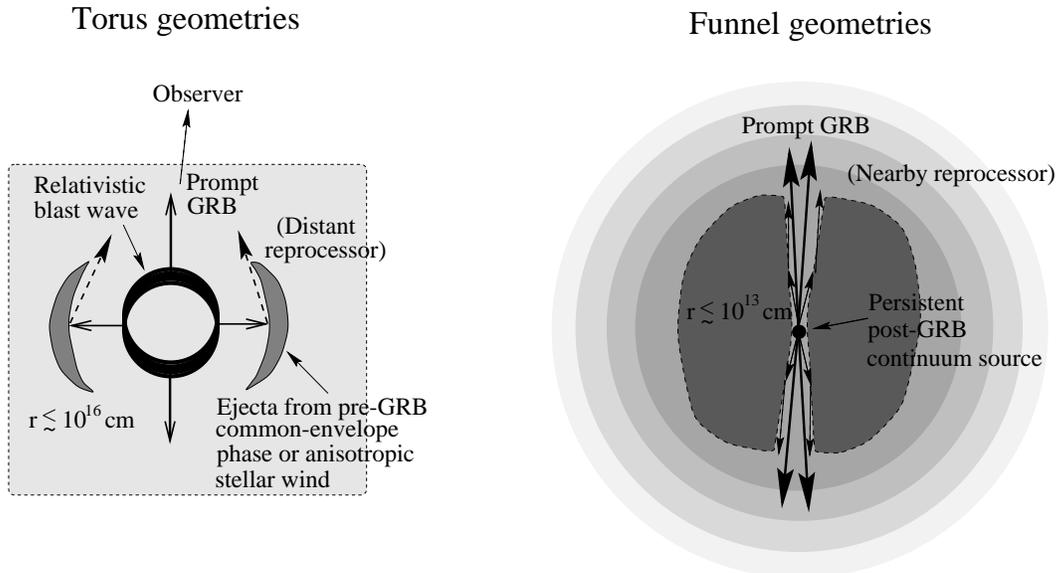}
\caption{Fundamental geometries of distant reprocessors (left)
and nearby reprocessors (right).}
\end{center}
\end{figure}

In a parameter-independent study, Lazzati et al. (2002) have calculated
the efficiency of reprocessing a given ionizing flux into fluorescence
and/or recombination line flux as a function of ionization parameter.
This efficiency is generally at most $\sim 1$~\% and dependent on the
precise spectral shape of the illuminating spectrum. This implies that
the observed iron lines listed in Tab. \ref{line_table} require a total
energy in the ionizing continuum of at least $\sim 10^{51}$~ergs (Ghisellini
et al. 2002). The reprocessing efficiency is drastically decreasing for
high ionization parameters $\xi \gtrsim 10^3$. Lazzati et al. (2002)
suggest that this may provide a diagnostic between distant and nearby
reprocessor models.

Another important aspect associated with iron K$\alpha$ line emission
is related to radioactive decay of ${^{56}{\rm Ni}} \to {^{56}{\rm Co}} 
\to {^{56}{\rm Fe}}$ in the ejecta of a supernova which may be associated 
with the GRB (Lazzati et al. 2001, McLaughlin et al. 2002, McLaughlin \& 
Wijers 2002). Since it is primarily $^{56}$Ni that is produced in the final
stages of a supernova progenitor, any emission line features should
gradually shift in energy from 8~keV for Nickel to 7.4~keV for Cobalt
and 6.9~keV for hydrogen-like iron on the radioactive decay time scales 
of 6.1~d and 78.8~d for $^{56}$Ni and $^{56}$Co, respectively. However, 
McLaughlin et al. (2002) and McLaughlin \& Wijers (2002) point out the
curious detail that $^{56}$Ni would usually predominantly decay via 
electron capture. Thus, if the reprocessing material happens to be
highly ionized, the decay time scale may be significantly delayed
w.r.t. the standard decay of neutral $^{56}$Ni. Consequently, for
detailed modeling of the ``iron'' line emission in GRB afterglows, 
a rest-frame line energy of $\sim 8$~keV might actually be a more 
appropriate assumption than the standard value of 6.4 -- 6.97~keV. 
In the remainder of this review, I will, for simplicity, continue 
to refer to any line feature around $E \sim 7$~keV in the GRB rest 
frame as ``iron line'', keeping in mind that it might have a 
significant contribution from Ni and Co.

In addition to light-travel time effects, standard constraints on 
reprocessor models are obtained considering the ionization and
recombination time scales. The ionization time scale may be estimated
as

\begin{equation}
t_{\rm ion} \approx {12 \over \int_{E_{\rm thr}}^{\infty} 
\Phi_{\rm ion} (E) \, \sigma_{\rm pi} (E) \, dE} \approx 
{12 \, (\alpha + 2) \over \sigma_0 \, \Phi_0 \, E_{\rm thr}},
\label{t_ion}
\end{equation}
where $E_{\rm thr}$ is the photoionization threshold energy,
$\sigma_{\rm pi} (E) \approx \sigma_0 \, (E / E_{\rm thr})^{-3}$
is the photoionization cross section, and $\Phi(E) = \Phi_0 \,
(E / E_{\rm thr})^{-\alpha}$ is the ionizing photon flux, assumed to 
be a straight power-law above the ionization threshold. Appropriate
averages for the various ionization stages of iron are $E_{\rm thr} 
\sim 8$~keV and $\sigma_0 \sim 3.5 \times 10^{20}$~cm$^{-2}$. The 
recombination time scale can be approximated as
\begin{equation}
t_{\rm rec} = {1 \over n_e \, \alpha_{\rm rec}} \approx {10^9 \, T_4^{3/4} 
\over n_H \, (Z_{\rm eff} / 26)^2} \; {\rm s},
\label{t_rec}
\end{equation}
where $\alpha_{\rm rec}$ is the recombination coefficient, $Z_{\rm eff}$
is the effective nuclear charge, $T_4$ is the electron temperature in 
units of $10^4$~K, and $n_H$ is the hydrogen number density in units
of cm$^{-3}$. The approximation for the recombination rate used above
assumes that recombination is dominated by radiative recombination,
which becomes inaccurate for ionization states lower than Fe~XXV at
temperatures above $\sim 10^5$~K. 

Additional model constraints are derived from considerations concerning 
optical-depth effects due to resonance scattering out of the line of 
sight (for the resonant Ly$\alpha$ line of Fe XXVI) and Thomson 
scattering, which define a maximum depth in the reprocessor beyond
which the material will effectively no longer contribute to the line
emission.

\subsection*{Distant Reprocessor Models}

Distant reprocessor models, in which the duration of the line
emission is dominated by the light travel time difference, 
$t_{\rm line} \sim R \, (1 + z) \, (1 - \cos\theta_{\rm obs}) / c$, were
the first to be discussed after it became clear that quasi-isotropic
fluorescence line emission scenarios were infeasible to explain the
observed iron line features (e.g., Lazzati et al. 1999, B\"ottcher
2000). Generally, in these models, a photoionizing continuum from 
the central source, emitted in tandem with the prompt and early
afterglow radiation, is impinging on a torus of dense, pre-ejected 
material. Recall, however, that the ionizing continuum does by no
means have to be identical or even similar to the observed GRB and
afterglow emission. The pre-ejected tori could plausibly be the result
of a common-envelope phase preceding the GRB event. All types of
GRB models pertaining to the class of black-hole accretion-disk
models, including the collapsar/hypernova and the He-merger (see, 
e.g., Fryer, Woosley \& Hartman for a review) are likely to have
undergone a common-envelope phase prior to the primary's core
collapse. The material ejected during such a common-envelope
phase is expected to have a directed velocity of the order of
the escape velocity from the secondary, $v \sim \sqrt{2 G M_{\rm sec} 
/ R_{\rm sec}} \sim 6 \times 10^7 \, (m/r)^{1/2}$~cm~s$^{-1}$, where
$M_{\rm sec} = m \, M_{\odot}$ and $R_{\rm sec} = r \, R_{\odot}$
are the secondary's mass and radius. For a detailed discussion of 
the expected structure of pre-ejected disks / tori, see B\"ottcher 
\& Fryer (2001) and references therein. In the collapsar/hypernova 
models, the delay between the ejection of the primary's hydrogen 
envelope and the GRB may be as large as 100,000~years, although 
significant uncertainties about the actual delay time scale remain. 
Such a long delay would place the pre-ejected material at radii
$R \gg 10^{16}$~cm, probably too large to be consistent with 
the observed time delays of the GRB X-ray emission line features. 

Much smaller delays are expected in the He-merger scenario (Zhang 
\& Fryer 2001) and the supranova model (Vietri \& Stella 1998).
In the He-merger model, time delays of a few hundred years to a 
few times 10,000 years may be typical, allowing for both nearby
and distant reprocessor scenarios. The supranova model predicts
delays of the order of the spin-down time scale of the supramassive
neutron star, $t_{\rm sd} \sim 10 \, j_{0.6} \, \omega_4^{-4} \, 
B_{12}^{-2}$~yr, where $j_{0.6}$ is the angular-momentum parameter
in units of 0.6, $\omega_4$ is the angular velocity in units of 
$10^4$~s$^{-1}$, and $B_{12}$ is the surface magnetic field in 
units of $10^{12}$~G. Since the typical magnetic field strength 
is very poorly constrained, delays of several months to several 
thousands of years might be possible for this model. The ejecta
velocity in this case should be more typical of supernova ejecta, 
$v \sim 10^9$~cm~s$^{-1}$. Thus it appears that the supranova model 
may be able to accomodate both distant and nearby reprocessor models 
as well as the thermal models discussed in the following section. 

Different geometrical variations of distant reprocessor models 
have been discussed in more detail by Lazzati et al. (2000) and 
Vietri et al. (2001). Analytical estimates as well as detailed 
numerical simulations of distant reprocessor scenarios (e.g., 
B\"ottcher 2000, Weth et al. 2000) seem to converge to a mass 
requirement of $M_{\rm fe} \sim 10^{-5}$ -- $10^{-4} \, M_{\odot}$ 
for most of the observed Fe~K$\alpha$ emission line features 
observed to date. Note, however, that Vietri et al. (2001) 
derive a significantly higher mass estimate of $M_{\rm Fe} 
\sim 1 \, M_{\odot}$ for GRB~991216, the most extreme case
in terms of total energy emitted in the Fe~K$\alpha$ line.

\subsection*{Nearby Reprocessor Models}

Nearby reprocessors, at typical distances of $\lesssim 10^{13}$~cm,
can either consist of pre-GRB ejecta (see discussion above) or the
expanding envelope of a super-giant GRB progenitor (e.g., Rees \&
M\'esz\'aros 2000, McLaughlin et al. 2002). In the latter case, if
the energy release in the GRB is strongly beamed, the envelope of
the progenitor star is expected to remain in a quasi-stable state
for a time of the order of the sound-crossing time through the 
progenitor, $t_{\rm sc} \sim 30 \, m_1^{-1/2} \, R_{13}^{3/2}$~d, 
where $m_1$ is the progenitor mass in units of $10 \, M_{\odot}$, 
and $R_{13}$ is its radius in units of $10^{13}$~cm. It can thus
plausibly survive the break-through of an ultrarelativistic jet,
and provide a scattering funnel for persistent ionizing radiation
throughout the prompt and early afterglow phase of the GRB. 

In this class of models, the highly collimated ultrarelativistic 
outflow, most probably associated with the prompt GRB and the
continuum afterglow emission, has moved far past the nearby
reprocessor and can obviously not contribute to the illuminating
continuum. Thus, these models require a persistent source of 
ionizing radiation over at least the duration over which the
iron line is observed. Rees \& M\'esz\'aros (2000) suggest 
that such a source could be provided by the gradually decaying 
energy flux from a magnetically driven relativistic wind from
a fast-rotating, strongly magnetized neutron star (magnetar), 
if the primary GRB mechanism does not result in the formation
of a black hole. They arrive at a required mass of $M_{\rm Fe}
\sim 10^{-8} \, M_{\odot}$ of iron in a very thin, ionized
skin of the funnel in order to reproduce an iron line with
properties as observed in GRB~991216. However, if a similar
abundance of iron is distributed throughout the entire stellar
envelope, then a mass estimate more in line with the distant
reprocessor models ($M_{\rm Fe} \sim 10^{-5}$ -- $10^{-4} \,
M_{\odot}$) results. 

These estimates could be reduced if the line emission originates 
in a dense medium with relativistic electron densities of $n_{e,
rel} \sim 10^{10}$ -- $10^{11}$~cm$^{-3}$. In such an environment, 
the line emission could be enhanced by the \v Cerenkov effect, as 
recently pointed out by Wang et al. (2002). Another variation of 
nearby reprocessor models has recently been proposed by Kumar \& 
Narayan (2002). They suggest the formation of a scattering 
screen at a distance of $\sim 10^{14}$ -- $10^{15}$~cm from the
GRB source, possibly as a result of $\gamma\gamma$ pair creation 
on back-scattered GRB radiation. GRB emission from later (internal)
shocks traveling along the jetted ejecta, are scattered back onto
the outside of the stellar envelope of the progenitor (the supernova
ejecta), and thus provide the ionizing flux for fluorescence line
emission. This model requires a smaller iron content in the ejecta
than previous models since it utilizes a larger fraction of the 
surface area of the progenitor's stellar envelope. Kumar \& Narayan
(2002) discuss this model in particular in light of the low-Z emission
lines in GRB~011211 (Reeves et al. 2002). In this specific case, 
their model would require a relatively large pre-GRB outflow rate,
producing a density in the surrounding medium of $n_0 \sim 7 \times 
10^7$~cm$^{-3}$, leading to an external-shock deceleration radius 
of $r_{\rm dec} \lesssim 5 \times 10^{14}$~cm. This would correspond 
to an observed deceleration time scale of $t_{\rm dec} \lesssim
0.8 \, \Gamma_2^{-2}$~s, which seems to be in conflict with the
duration of $t_{\rm dur} \sim 270$~s of GRB~011211, with no indication
of rapid variability on the time scale of the order of $t_{\rm dec}$.

\section*{THERMAL MODELS}

As an alternative to photoionization models, Vietri et al. (1999)
had suggested a thermal model in the framework of the supranova
model. They argue that a relativistic fireball associated with the
GRB might hit the pre-GRB supernova remnant within $\sim 10^3$~s
and heat the ejecta to $T \sim 3 \times 10^7$~K. At such temperatures,
the plasma emission is expected to show strong thermal bremsstrahlung
emission as well as line emission, in particular strong Fe~K$\alpha$
recombination line emission. They suggest that the bremsstrahlung and 
recombination continuum may explain the secondary X-ray outburst
observed in GRB~970508. Since the supranova model seems to be consistent
with a large range of SN -- GRB delays, one might expect that secondary
X-ray outbursts and delayed X-ray emission line features on a variety
of time scales can be explained with this type of models. General
constraints on thermal emission scenarios for Fe~K$\alpha$ lines
have also been considered by Lazzati et al. (1999).

B\"ottcher \& Fryer have investigated the thermal X-ray emission from
shock-heated pre-ejected material in alternative progenitor models
(i.e. other than the supranova model), such as the collapsar/hypernova 
and the He-merger model. They found that the He-merger scenario 
provides a feasible setting for the production of transient Fe~K$\alpha$
line emission in the range of luminosities and durations observed. 
Since both the He-merger and the hypernova/collapsar models are also
consistent with much larger SN -- GRB delays, they predict that many
GRBs may display late-time, thermal X-ray flashes from the shock-heating
of pre-ejected material from a common-envelope phase, which could be
detected out to redshifts of $\sim 1$ with currently operating X-ray
telescopes.

\begin{figure}
\begin{center}
\includegraphics[width=13cm]{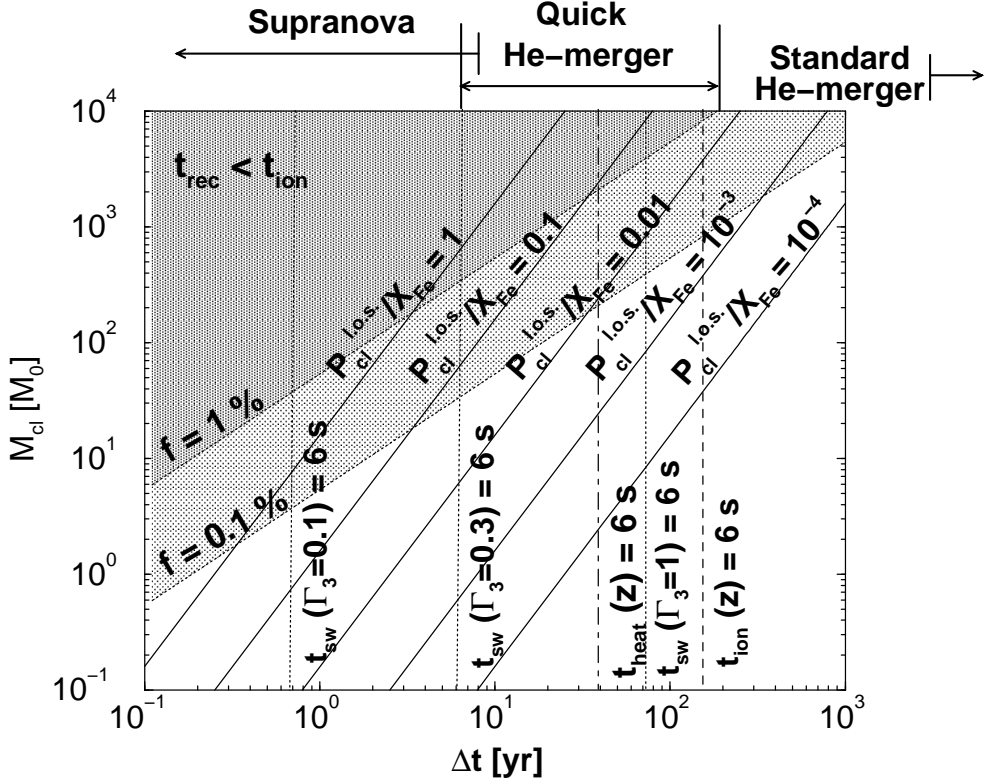}
\caption{Parameter constraints concerning supernova ejecta mass
$M_{\rm cl}$ concentrated in dense clumps, and time delay $\Delta t$
between the primary's supernova explosion and the GRB. Solid lines 
indicate the condition that an Fe~K absorption edge of the depth
observed in GRB~990705 is produced for various values of the ratio
of the probability $P_{\rm cl}^{\rm l.o.s.}$ of an absorbing cloud
being located in the line of sight, to the iron enhancement, $X_{\rm Fe}$,
with respect to standard solar system values. Constellations which would
give a consistent physical scenario must either be located close to
the vertical line corresponding to $t_{\rm ion} (z) = 6$~s (if
recombination is inefficient) or within the shaded regions in the
upper left corner of the plot, which indicates the condition $t_{\rm rec}
\le t_{\rm ion}$ for volume-filling factors of the SN ejecta of
1~\% and 0.1~\%, respectively, 1~year after the SN.}
\end{center}
\label{abs_parameters}
\end{figure}

An alternative scenario based on thermal emission has recently been
suggested by M\'esz\'aros \& Rees (2001). As the collimated outflow
from the central engine of a collapsar is piercing through the stellar
envelope of the progenitor, a substantial amount of energy is deposited
into the stellar material, which might be highly magnetized. After the
jet breaks out of the stellar envelope, this plasma bubble becomes buoyant
and emerges through the evacuated funnel of the envelope within $\sim
10^4$ -- $10^5$~s. At that time, it may have attained a temperature of
$\sim 10^6$ -- $10^7$~K, sufficient to produce iron lines, but potentially
also a plasma emission spectrum dominated by lower-energy lines, as
possibly observed in GRB~011211. 

\section*{MODEL IMPLICATIONS OF THE TRANSIENT ABSORPTION FEATURE}

In principle, inferring constraints on parameters of the circumburster
material from observed GRB properties is an easier task than inferring
them from emission lines, because in the case of absorption features
the continuum responsible for photoionization is identical to the
observed GRB and afterglow continuum. Time-dependent X-ray absorption 
features had been studied for generic, quasi-homogeneous environments 
by B\"ottcher et al. (1999) and Ghisellini et al. (1999), and for more 
general cases, including radial gradients, by Lazzati et al. (2001) and 
Lazzati \& Perna (2002). For given values of the depth $\tau_{\rm edge}$ 
of an absorption edge and the time scale $t_{\rm edge}$ within which 
it is disappearing, one can directly infer a characteristic radius 
(by setting $t_{\rm ion} = t_{\rm edge}$). Combined with the column
density derived from $\tau_{\rm line}$, this allows a direct estimate
of the (isotropic) amount of iron in the absorber. 

It came as a big surprise that these estimates, applied to the parameters 
of the transient absorption line in GRB~990705 (Amati et al. 2000), 
yielded an estimate of $M_{\rm Fe} \sim 44 \, \Omega \, M_{\odot}$
within $R \lesssim 1.3$~pc, where $\Omega$ is the solid angle covered 
by the absorber as seen from the GRB source. Since this does obviously 
not seem realistic in any known astrophysical setting, additional effects
due to clumping of the absorber in small clouds, in which recombination
would become efficient (e.g., B\"ottcher et al. 2001), or resonance
scattering of the Fe~XXVI~Ly$\alpha$ line out of the line of sight
(Lazzati et al. 2001) had been considered. Both of these effects 
could plausibly reduce the necessary amount of iron in the absorber
to $M_{\rm Fe} \lesssim 1 \, M_{\odot}$, but require a rather extreme
degree of clumping, with densities of $n \sim 10^{11}$~cm$^{-3}$ in
the clumps and distance/size ratios of $x/r \sim 10^3$ -- $10^5$. 

B\"ottcher et al. (2002) have scaled the required parameters of the 
absorber to the clumping properties of supernova ejecta, derived from 
detailed 3-D hydrodynamics simulations of supernovae. The results
were parameterized in terms of the total mass contained in the dense
absorbing clouds, $M_{\rm cl}$, and the time delay $\Delta t$ between 
the supernova producing the absorbing ejecta, and the GRB. In the
case of a supranova scenario, $\Delta t$ is the delay between the
progenitor supernova and the GRB, while in the He-merger scenario,
$\Delta t$ represents the time between the primary's supernova
explosion and the He-merger-triggered GRB. Other progenitor models,
such as the collapsar/hypernova models, would possess too dilute
environments in order to be consistent with the observed properties
of GRB~990705. The results of B\"ottcher et al. (2002) are summarized 
in Fig. 2, and illustrate that all currently discussed GRB models seem
to be hard-pressed in order to produce the required environments to
reproduce the transient absorption feature in GRB~990705, with the
possible exception of the supranova scenario.

\end{document}